\documentclass[10pt,twocolumn,twoside,journal]{IEEEtran}
\usepackage{algorithm,algorithmic,amsmath,amssymb,epsfig,enumerate,amsfonts,url,multirow,array,tabularx,booktabs,cite,stfloats,flushend}
\usepackage[T1]{fontenc}
\usepackage{threeparttable}
\newcommand{\bm}[1]{\mbox{\boldmath{$#1$}}}

\usepackage{bigstrut, multirow, rotating}

\def\x{{\mathbf x}}

\def\y{{\mathbf y}}

\def\z{{\mathbf z}}

\def\f{{\mathbf f}}
\def\m{{\mathbf m}}

\def\A{{\mathbf A}}

\def\Y{{\mathbf Y}}
\def\X{{\mathbf X}}
\def\T{{\mathbf T}}
\def\N{{\mathbf N}}

\DeclareMathAlphabet{\mathcal}{OMS}{cmsy}{m}{n}

\newcolumntype{I}{!{\vrule width 3pt}}
\newlength\savedwidth

\newlength\savewidth

\ifCLASSINFOpdf

\else

\fi

\begin{document}

\title{Linear Regression for Speaker Verification}

\author{~Xiao-Lei~Zhang

\thanks{
Xiao-Lei Zhang is with the Center for Intelligent Acoustics and Immersive Communications, School of Marine Science and Technology, Northwestern Polytechnical University, Xi'an, China (e-mail: xiaolei.zhang9@gmail.com). }
}


\maketitle

\begin{abstract}
This paper presents a linear regression based back-end for speaker verification. Linear regression is a simple linear model that minimizes the mean squared estimation error between the target and its estimate with a closed form solution, where the target is defined as the ground-truth indicator vectors of utterances. We use the linear regression model to learn speaker models from a front-end, and verify the similarity of two speaker models by a cosine similarity scoring classifier. To evaluate the effectiveness of the linear regression model, we construct three speaker verification systems that use the Gaussian mixture model and identity-vector (GMM/i-vector) front-end, deep neural network and i-vector (DNN/i-vector) front-end, and deep vector (d-vector) front-end as their front-ends, respectively. Our empirical comparison results on the NIST speaker recognition evaluation data sets show that the proposed method outperforms within-class covariance normalization, linear discriminant analysis, and probabilistic linear discriminant analysis, given any of the three front-ends.
\end{abstract}

\IEEEpeerreviewmaketitle

 \setlength{\arraycolsep}{0.0em}

\section{Introduction}\label{sec:introduction}
\bstctlcite{IEEEexample:BSTcontrol}

\IEEEPARstart{S}{peaker} verification has long been a fundamental task in speech processing. A speaker verification system verifies an identity claim made by a test speaker, and decides to accept or reject the claim. It can be either {text-dependent} or {text-independent} based on its input speech materials: the former constrains a speaker to pronounce a prescribed text, while the latter does not constrain the speech contents. This paper studies \textit{text-independent} speaker verification. A text-indepdent speaker verification system generally contains two components. The first component is a front-end which extracts a feature vector from a speaker utterance by some density estimator. The second component is a back-end which builds speaker models and measures the similarity of two speaker models by a classifier.

An early speaker verification front-end is feature averaging which learns a feature vector from a speaker utterance by averaging the frame-level acoustic features \cite{markel1977long}. The method requires long speech utterances to reach stable speech statistics. Another class of front-ends is statistical models, which estimates the density of speech frames by statistical models. Early approaches of this kind build a model, e.g. vector quantization \cite{soong1987report} or Gaussian mixture model (GMM) \cite{reynolds1995speaker,reynolds1995robust}, for each speaker. These approaches are inefficient when the number of speakers is large. To alleviate this problem, in \cite{reynolds2000speaker}, Reynolds \textit{et al.} proposed a GMM-based universal background model (GMM-UBM) which builds a single GMM from the pool of all training speakers. GMM-UBM is a fundamental method of speaker verification in recent years. To deal with speaker and channel variability, many approaches were proposed along with GMM-UBM, where factor analysis \cite{dehak2011front} is among the effective ones. It first extracts high-dimensional supervectors of utterances which are their first- and second-order statistics produced from GMM-UBM, and then reduces the supervectors to low-dimensional \textit{identity vectors} (i-vectors) by factor analysis. The above combination of GMM-UBM and i-vector is the GMM/i-vector front-end.

 Recently, deep neural network (DNN) based front-ends have received much attention \cite{sarkar2014combination,variani2014deep,lei2014novel}. In  \cite{sarkar2014combination}, Sarkar \textit{et al.} used DNN to extract frame-level bottleneck features that were then used as the input of GMM-UBM. In  \cite{lei2014novel}, Lei \textit{et al.} took a DNN trained for a different task, e.g. speech recognition, to generate posterior probability of speech frames, which is a supervised alternative to GMM-UBM, and then used the factor analysis \cite{dehak2011front} to extract i-vectors from the DNN based UBM. The method is denoted as the DNN/i-vector front-end. To demonstrate the advantages of the DNN/i-vector front-end, its DNN acoustic model needs to be trained with additional data \cite{richardson2015deep}. In \cite{variani2014deep}, Variani \textit{et al.} trained a DNN classifier
to map frame-level features in a given context to the corresponding speaker identity target, and extracted a feature vector, referred as a deep vector or ``d-vector'', from a speaker utterance by averaging the activations derived from the last DNN
hidden layer. The method is known as the d-vector front-end.

After feature extraction by a front-end, a speaker verification back-end builds speaker models for classification. It generally contains two stages---a development stage and a test stage. The development stage builds a \textit{speaker space} from development data, where each speaker acts like a coordinate axis of the space. The test stage gets the enrollment and test speaker models of a trial from the speaker space and then evaluates the similarity of the two models by a classifier.

 We summarize some back-ends as follows. In \cite{reynolds2000speaker}, Reynolds \textit{et al. } first built a speaker space by adapting the GMM-UBM to many speaker-dependent GMMs by the maximum a posteriori estimation in the development stage and then verifies the identity of a test speaker by a likelihood ratio test. Later on, in \cite{campbell2006support}, Campbell \textit{et al.} trained a support vector machine classifier to distinguish true speakers from imposter speakers with nuisance attribute projection \cite{campbell2006svm} for compensating session variability. In \cite{dehak2011front}, Dehak \textit{et al.} proposed to learn a speaker space by within class covariance normalization (WCCN) or linear discriminant analysis (LDA) and then applied cosine similarity scoring as the classifier. In \cite{kenny2010bayesian}, Kenny proposed to extract speaker models from an i-vector based front-end or LDA and then used probabilistic LDA (PLDA) as the classifier. Besides, in \cite{snyder2016deep}, Snyder \textit{et al.} proposed an end-to-end training method to train a DNN based front-end and a PLDA-like back-end jointly.


In this paper, we propose a linear regression (LR) based back-end. LR is a traditional statistical regression model that minimizes the mean squared error between the target and its estimate with a closed-form solution. In the development stage of the back-end, we apply LR to learn a speaker space where the target of the LR model is the ground-truth indicator vectors of the speaker utterances. In the enrollment and test stages, we first extract the enrollment and test speaker models of a trial from the speaker space, and then evaluate the similarity of the two models by cosine similarity scoring. The overall back-end is denoted as the LR+cosine back-end. To evaluate its effectiveness, we propose three speaker verification systems which combine the LR+cosine back-end with the GMM/i-vector, DNN/i-vector, and d-vector front-ends, respectively.

We have conducted an extensive experiment on the NIST 2006 speaker recognition evaluation (SRE) and NIST 2008 SRE data sets. We have compared the LR+cosine back-end with the cosine similarity scoring (cosine), WCCN with the cosine similarity scoring (WCCN+cosine), LDA with the cosine similarity scoring (LDA+cosine), and LDA with the PLDA scoring (LDA+PLDA) back-ends. Our experimental results show that the proposed method outperforms the comparison methods, and the experimental conclusion is consistent in different lengths of enrollment speech.

This paper is organized as follows. In Section \ref{sec:format}, we introduce the LR-based back-end and three LR-based speaker verification systems. In Section \ref{sec:typestyle}, we present the experiments. In Section \ref{sec:conclusion}, we summarize the paper.




%
%

\section{Linear regression for speaker verification}
\label{sec:format}
In this section, we first present the LR-based back-end in Section \ref{subsec:lr}, and then present three front-ends that will be combined respectively with the LR-based back-end in Section \ref{subsec:front_end}.

 The procedure of any of the three speaker verification systems is as follows. The front-end extracts a feature vector $\x$ from an utterance $\{\z_k\}_{k=1}^s$ where $s$ denotes the number of frames of the utterance. Then, the LR-based back-end first gets the speaker model $\m$ from $\x$ by the LR model and then verifies the identity of ${\m}$ by the cosine similarity scoring.


\subsection{Linear regression based back-end}\label{subsec:lr}

Suppose a labeled development corpus after processed by a front-end is given by $\{\left\{(\x_{i,j}, y_{i,j})\right\}_{j=1}^{U_i}\}_{i=1}^S$ where $S$ is the number of speakers, $U_i$ is the number of utterances of the $i$-th speaker, $\x_{i,j}$ is the feature vector of a speaker utterance produced from a front-end, and $y_{i,j}$ is the ground-truth label of the utterance representing the identification of the speaker, $1\leq y_{i,j}\leq S$. Suppose $y_{i,j}=k$, then we change $y_{i,j}$ to an $S$-dimensional indicator vector $\y_{i,j}$ which is a binary code with the $k$-th dimension set to $1$ and the other dimensions set to $0$. As a result, we can rewrite the labeled corpus as $\{\left\{(\x_{i,j}, \y_{i,j})\right\}_{j=1}^{U_i}\}_{i=1}^S$. We fit $\{\left\{(\x_{i,j}, \y_{i,j})\right\}_{j=1}^{U_i}\}_{i=1}^S$ to a LR model:
   \begin{eqnarray}
\y_{i,j} = \mathbf{A}^T\x_{i,j}+\bm\epsilon_{i,j}
\label{eq:1}
 \end{eqnarray}
 where $\A$ is the LR model and $\epsilon_{i,j}$ is the estimation error. Minimizing the squared estimation error $\|\bm\epsilon\|^2_2$ derives the following closed-form solution:
    \begin{eqnarray}
\A = (\X\X^T)^{-1}\X\Y^T
\label{eq:2}
 \end{eqnarray}
 where
    \begin{eqnarray}
&&\X=[\x_{1,1},\ldots,\x_{1,U_1},\ldots,\x_{i,j},\ldots,\x_{S,1},\ldots,\x_{S,U_S}],\nonumber\\
&&\Y=[\y_{1,1},\ldots,\y_{1,U_1},\ldots,\y_{i,j},\ldots,\y_{S,1},\ldots,\y_{S,U_S}].\nonumber
\label{eq:3}
 \end{eqnarray}

In the enrollment and test stages, we apply the LR model to extract a new feature $\hat{\y}$ from $\x$ by the following equation:
   \begin{eqnarray}
\hat{\y} = \mathbf{A}^T\x.
\label{eq:4}
 \end{eqnarray}
 The speaker model $\m$ is given by:
    \begin{eqnarray}
\m = \frac{1}{V}\sum_{v=1}^V\hat{\y}_v
\label{eq:4}
 \end{eqnarray}
where $V$ is the number of utterances of the speaker.

Finally, we employ a classifier to identify the similarity of two speaker models ${\m}^{\footnotesize\mbox{enroll}}$ and ${\m}^{\footnotesize\mbox{test}}$. Despite that many classifiers could be applied, we use the simple and effective cosine similarity scoring as an example based on the experimental conclusion of reference \cite{dehak2011front}. The cosine similarity of the two models is calculated by:
  \begin{eqnarray}
{\rm{score}}({\m}^{\footnotesize\mbox{enroll}}, {\m}^{\footnotesize\mbox{test}})=\frac{\left<{\m}^{\footnotesize\mbox{enroll}},{\m}^{\footnotesize\mbox{test}}\right>}{\|{\m}^{\footnotesize\mbox{enroll}}\|\mbox{ }\|{\m}^{\footnotesize\mbox{test}}\|}\gtreqless \theta
\label{eq:cosine}
 \end{eqnarray}
which is compared with a decision threshold $\theta$. If the score is larger than $\theta$, then the two models are judged as from the same speaker; otherwise, they are from different speakers.

\subsection{Front-ends}\label{subsec:front_end}

The three speaker verification systems based on the LR-based back-end use the GMM/i-vector \cite{reynolds2000speaker,dehak2011front}, DNN/i-vector \cite{lei2014novel}, and d-vector \cite{variani2014deep} front-ends, respectively.

\subsubsection{GMM/i-vector front-end}
The GMM/i-vector front-end contains a GMM-UBM $\Omega$ \cite{reynolds2000speaker,kenny2008study} which is a speaker- and channel-independent GMM trained from the pool of all speech frames of the development data, and a total variability matrix $\T$ \cite{dehak2011front} that encompasses both
speaker- and channel-variability. Suppose $\Omega$ contains $C$ Gaussian mixture components, and suppose we have an utterance of $L$ frames $\{\z_l\}_{l=1}^L$ where $\z_l$ is a $F$-dimensional acoustic feature. The zero-th order and centralized first-order Baum-Welch statistics of the utterance extracted from the $c$-th component of $\Omega$ is:
   \begin{eqnarray}
&&n_c = \sum_{l=1}^L {\rm{P}}(c|\z_l, \Omega),\\
&&\f_c =  \sum_{l=1}^L {\rm{P}}(c|\z_l, \Omega)(\z_l-\bm\mu_c)
\label{eq:GMM_1}
 \end{eqnarray}
 where $\bm\mu_c$ is the mean of the $c$-th component of $\Omega$. If we define $\N$ as a $CF\times CF$-dimensional diagonal matrix whose diagonal blocks are $n_c\mathbf{I}$, $\bar{\f}=[\f_1^T,\ldots,\f_C]^T$ as a supervector, and $\bm\Sigma$ as a $CF\times CF$-dimensional diagonal covariance matrix  estimated during factor analysis training \cite{dehak2011front}, then we obtain the i-vector $\x$ by:
    \begin{eqnarray}
\x = (\mathbf{I}+\T^T\bm\Sigma^{-1}\N\T)^{-1}\T^T\bm\Sigma^{-1}\bar{\f}
\label{eq:GMM_2}
 \end{eqnarray}
 where $\T$ and $\bm\Sigma$ is invariant across utterances.

\subsubsection{DNN/i-vector front-end}
The difference between the DNN/i-vector front-end \cite{lei2014novel} and the GMM/i-vector front-end is that the GMM-UBM in the DNN/i-vector front-end is estimated by a DNN acoustic model trained for automatic speech recognition. Specifically, the DNN acoustic model is used to estimate the senone posteriors of acoustic features, where a senone is used to model the tied states of a set of triphones that are close in the acoustic space. If we model the posterior distribution of a senone by a Gaussian mixture component of the GMM-UBM, then we can use the senone posteriors to train the GMM-UBM in the following way.

Suppose the development corpus contains $U$ utterances, and the $u$-th utterance has $L_u$ frames $\{\z^{(u)}_{l}\}_{l=1}^{L_u}$. The parameters of the GMM-UBM are estimated by:
    \begin{eqnarray}
&&\gamma^{(u)}_{c,l} \approx {\rm{P}}(c|\z^{(u)}_l), \\
&&\pi_{c} = \sum_{u=1}^U\sum_{l=1}^{L_u} \gamma^{(u)}_{c,l},\\
&&\bm\mu_c = \frac{\sum_{u=1}^U\sum_{i=1}^{L_u}\gamma^{(u)}_{c,l}\z^{(u)}_l  }{\sum_{u=1}^U\sum_{i=1}^{L_u}\gamma^{(u)}_{c,l}},\\
&&\bm\Sigma_c = \frac{ \sum_{u=1}^U\sum_{i=1}^{L_u}\gamma^{(u)}_{c,l}\z^{(u)}_l{\z^{(u)}_l}^T  }{  \sum_{u=1}^U\sum_{i=1}^{L_u}\gamma^{(u)}_{c,l} } -\bm\mu_c \bm\mu_c^T
 \end{eqnarray}
 where $\{\gamma^{(u)}_{c,l}\}_{c=1}^C$ represent the alignments of $\z^{(u)}_l$ which are the posteriors of $\z^{(u)}_l$ produced by the DNN acoustic model, $\pi_{c}$ and $\bm\Sigma_c$  are the prior and covariance of the $c$-th mixture component, respectively.

 The DNN acoustic model is trained in a supervised mode, where the ground-truth labels of the speech frames are the alignments produced by a hidden-Markov-model-GMM (HMM-GMM) speech recognition system. It usually adopts a contextual window with a window size of, e.g. $(2W+1)$, to expand the input from $\z_l$ to $[\z_{l-W}^T,\ldots,\z_l^T,\ldots, \z_{l+W}^T]^T$ where $W$ is the half-window length.

\subsubsection{D-vector front-end}
The d-vector front-end \cite{variani2014deep} averages the frame-level features of an utterance produced from the top \textit{hidden} layer of a DNN classifier for an utterance-level d-vector $\x$. The DNN is trained to minimize the classification error of speech frames, where the ground-truth label of a speech frame is the indicator vector $\y$ of the speaker that the speech frame belongs to. The DNN adopts a \textit{large} contextual window with a window size of $(2W_d+1)$ to expand its input acoustic feature from $\z_l$ to $[\z_{l-W_d}^T,\ldots,\z_l^T,\ldots, \z_{l+W_d}^T]^T$, which is important in improving the effectiveness and robustness of the d-vector front-end.


\section{Experiments}
\label{sec:typestyle}

In this section, we present the databases and evaluation metrics at first in Section \ref{subsec:3_1}, then the experimental setup in Section \ref{subsec:3_2}, and finally the experimental results in Sections \ref{subsec:result} and \ref{subsec:3_4}.

\subsection{Databases and evaluation metrics}\label{subsec:3_1}

We took the 8\textit{conv} condition of NIST 2006 speaker recognition evaluation (SRE) database as the development set, and the 8\textit{conv} condition of NIST 2008 SRE for enrollment and test. The 8\textit{conv} condition of NIST 2006 SRE contains 402 female speakers and 298 male speakers. The 8\textit{conv} condition of NIST 2008 SRE contains 395 female speakers and 240 male speakers. Each speaker has 8 conversations. A speaker utterance in a conversation was about 1 to 2 minutes long after removing the silence segments by VAD, where we took its ASR transcript as its VAD label. We split all speech signals into 15 second segments.

To illustrate the global performance of the proposed method in terms of detection error tradeoff (DET) curves, we built an \textit{initial} test condition as follows. We selected the first 150 second speech of the first conversation of a speaker as the enrollment data of the speaker, and split the last 30 second speech of the 6-th conversation of the speaker into two test segments with each segment as an individual test. We took each speaker as a claimant with the remaining speakers acting as imposters, and rotated through the tests of all speakers. We conducted the experiment on females and males respectively. The number of claimant and imposter trials are summarized in Table \ref{tab:data}. The closer the DET curve approaches to the origin, the better the performance is.

To investigate how the performance of the proposed method varies with the length of the enrollment speech, we conducted experiments in six test conditions described in Table \ref{tab:scenario}. Specifically, for each speaker in the 8\textit{conv} condition of the NIST 2008 SRE, we first randomly picked 2 segments from a randomly selected conversation with each segment as an individual test; then, we randomly selected $X$ segments from the remaining 7 conversations as the enrollment data of the speaker, where we set $X$ to 1, 2, 3, 5, 10, and 15 for the six test conditions respectively. For a given test condition, we built the claimant and imposter trials in the same way as the initial test condition. Therefore, the number of the trials are the same as that in Table \ref{tab:data}. Because the enrollment and test speech of a trial was selected randomly, we ran the experiments on each test condition 100 times and reported the average results so as to prevent biased conclusions.
We used equal error rate (EER), minimum detection cost function (DCF) with SRE'08 parameters (DCF$\scriptsize{\mbox{08}}$), and minimum DCF with SRE'10 parameters (DCF$\scriptsize{\mbox{10}}$) as the evaluation metrics. The smaller the EER or DCF is, the better the performance is.

 \begin{table} [t]
\caption{\label{tab:scenario} {Description of test conditions.}}
\centerline{
\scalebox{1}{
\begin{tabular}{ccc}
\hline
Name & Length of enrollment speech & Length of test speech  \\
 \hline
15"-15" & 15 seconds & 15 seconds\\
30"-15" & 30 seconds & 15 seconds\\
45"-15" & 45 seconds & 15 seconds\\
75"-15" & 75 seconds & 15 seconds\\
150"-15" & 150 seconds & 15 seconds\\
225"-15" & 225 seconds & 15 seconds\\
\hline
\end{tabular}}
}
\end{table}

 \begin{table} [t]
\caption{\label{tab:data} {Number of claimant and imposter trials.}}
\centerline{
\scalebox{1}{
\begin{tabular}{lccc}
\hline
 & \#speakers & \#true trials  & \#imposter trials  \\
 \hline
Female& 395 & 790 & 311,260 \\
Male& 240 & 480 & 114,720\\
\hline
\end{tabular}}
}
\end{table}

\subsection{Experimental setup}\label{subsec:3_2}

\subsubsection{Acoustic features}
We set the frame length to 25 ms and the frame shift to 10 ms. We extracted 19-dimensional mel-frequency cepstral coefficients (MFCC), 13-dimensional relative spectral filtered perceptual linear predictive cepstral coefficients (RASTA-PLP) and 1-dimensional log energy, as well as their delta and delta-delta coefficients from each frame, which produced a total of 99-dimensional acoustic feature per frame.

\subsubsection{Front-ends}
For the GMM/i-vector front-end, we used gender-dependent UBMs containing 2048 Gaussian mixtures and 400 total factors defined by the total variability matrix $\T$. We followed  the MSR identity toolbox for the implementation of the GMM/i-vector front-end.

For the DNN/i-vector front-end, we trained a DNN acoustic model from the Switchboard-1 database. The alignments of the frames for the DNN training, which contained 8730 senones, were generated by a HMM-GMM speech recognition system implemented in the Kaldi pipeline. The half-window length $W$ of the DNN was set to 3, which expanded the acoustic features to 693 dimensions. As a result, the DNN acoustic model used the 693-dimensional feature as the input and its corresponding 8730 dimensional alignment as the ground-truth label. The DNN has $7$ hidden layers, each of which consists of $2048$ rectified linear units. The output units of the DNN are the softmax functions. The DNN was optimized by the minimum cross-entropy criterion. The number of epoches for backpropagation training was set to $50$. The batch size was set to $512$. The learning rate of the stochastic gradient descent was set to $0.1$. The momentum was set to 0.5 for the first 10 epoches, and set to 0.9 for the other epoches. The dropout rate of the hidden units was set to 0.2. We used the posterior probability of the development data produced by the DNN acoustic model to train gender-dependent UBMs. Because many senones have small posterior probabilities, we truncated the UBMs from 8730 Gaussian mixtures to 3096 Gaussian mixtures by discarding the mixtures that have small zero-th order Baum-Welsh statistics. We used 400 total factors to generate the i-vectors.

For the d-vector front-end, we trained gender-dependent DNNs on the development data, where the two DNNs have the same parameter setting as follows. The half-window length $W_d$ was set to 20, which expanded the acoustic feature to 4059 dimensions. Each DNN has $4$ hidden layers, each of which consists of 400 rectified linear units. The output dimensions of the two DNNs are 395 for the females and 240 for the males, respectively. The learning rate of the stochastic gradient descent was set to $0.008$. All other parameters were set to the same values as those in the DNN/i-vector front-end.

\subsubsection{Back-ends in comparison}
We compared the LR+cosine back-end with the following back-ends:
\begin{itemize}
  \item \textbf{Cosine similarity scoring (cosine):} The cosine back-end evaluates the cosine similarity of two speaker models directly where the speaker model is simply an average of the utterance-level feature vectors of the speaker produced from a front-end  \cite{dehak2011front}.
  \item \textbf{WCCN+cosine:} WCCN helps compensate for channel variability \cite{hatch2006within}. It was first proposed for a SVM based back-end, and then applied to the cosine similarity scoring by Dehak \textit{et al.} \cite{dehak2011front}. Here we compared with the WCCN+cosine method \cite{dehak2011front}.
  \item \textbf{LDA+cosine:} LDA is a supervised dimensionality reduction method. Dehak \textit{et al.} \cite{dehak2011front} applied LDA to the cosine similarity scoring. Here we set the output dimension of LDA to 200 in all evaluations, which is a common experimental setting in literature.
  \item \textbf{LDA+PLDA:} The PLDA classifier was first introduced to speaker verification by Kenny in \cite{kenny2010bayesian}. LDA is usually used as a feature extractor for PLDA. We set the output dimension of LDA to 200 in all evaluations.
\end{itemize}


\subsection{Results}\label{subsec:result}

\begin{figure}
 \centering
         \includegraphics[width=7.5cm]{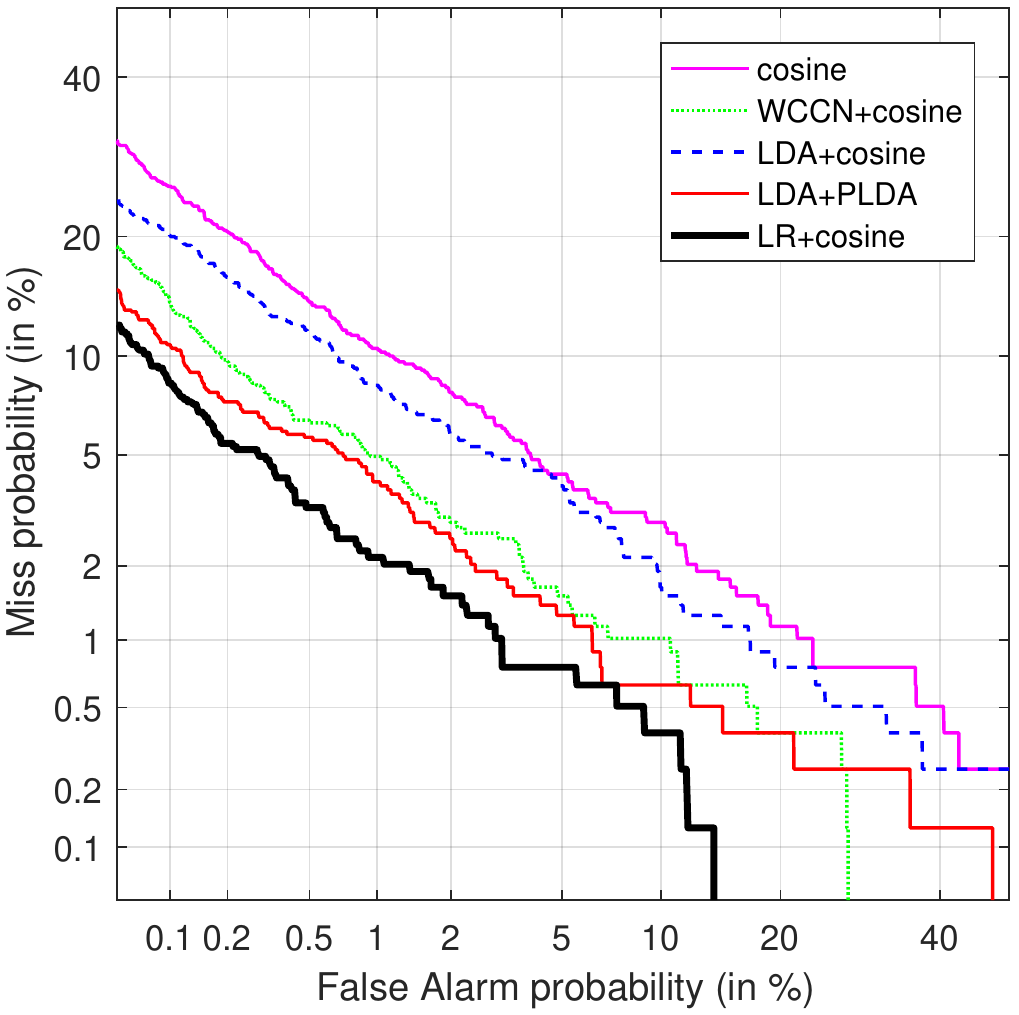}
         \caption{DET curves of the back-ends with the GMM/i-vector front-end on the female part of the initial test condition.}
 \label{fig:det_GMM_ivector_female}
 \end{figure}

\begin{figure}
 \centering
         \includegraphics[width=7.5cm]{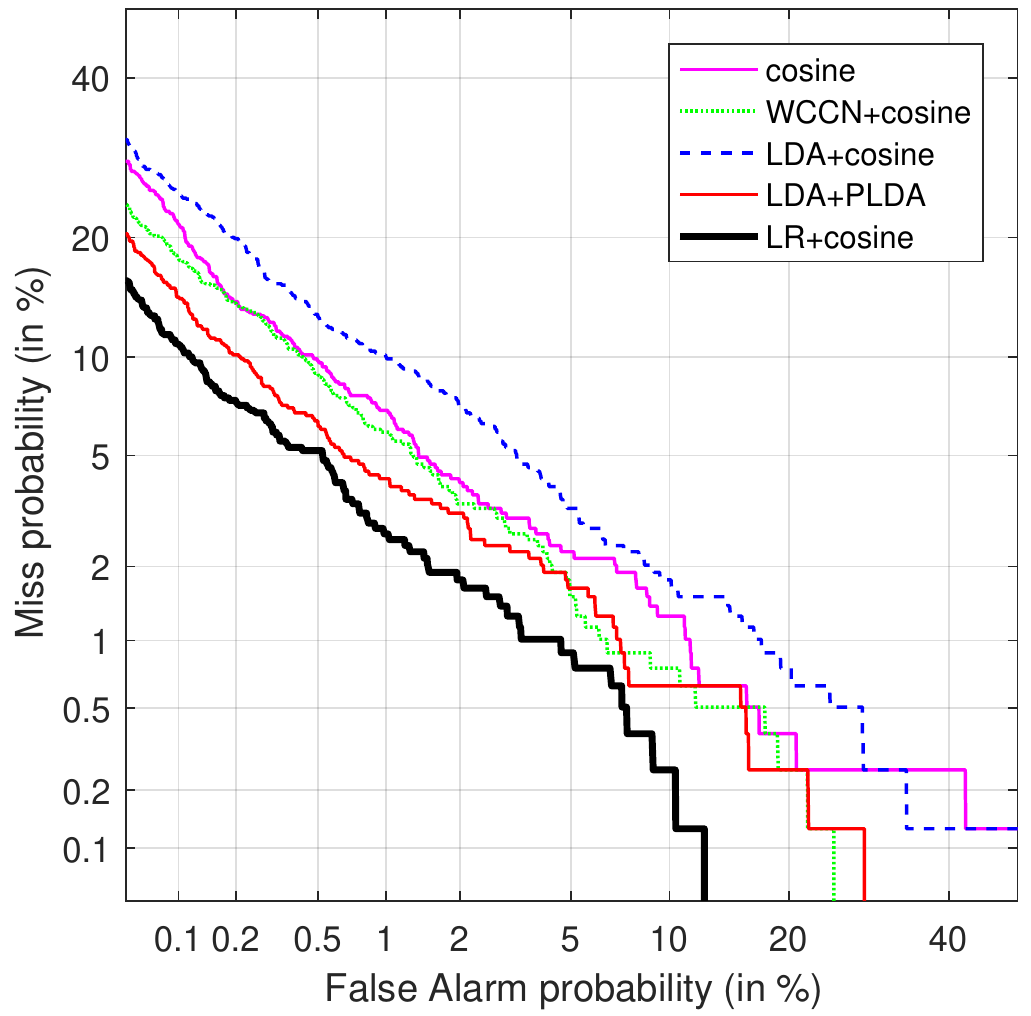}
         \caption{DET curves of the back-ends with the DNN/i-vector front-end on the female part of the initial test condition.}
 \label{fig:det_DNN_ivector_female}
 \end{figure}

\begin{figure}
 \centering
         \includegraphics[width=7.5cm]{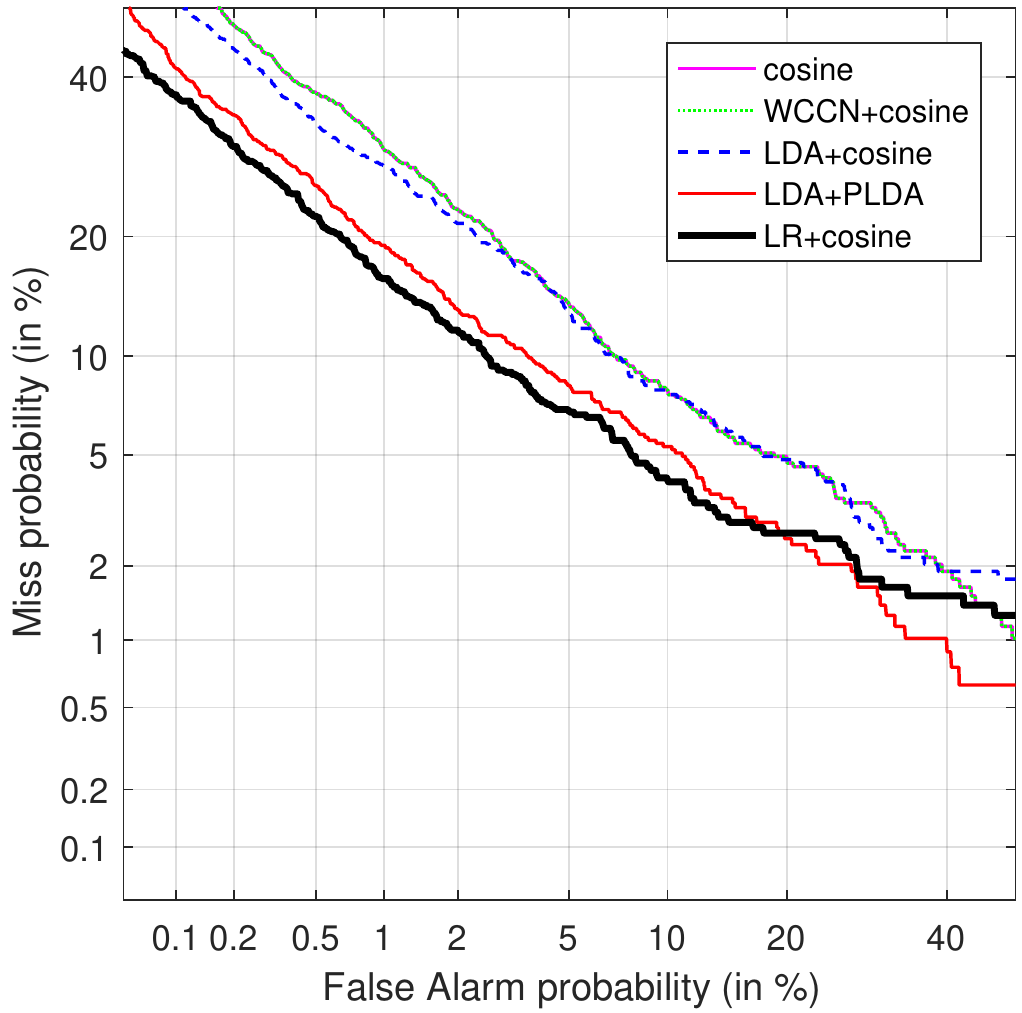}
         \caption{DET curves of the back-ends with the d-vector front-end on the female part of the initial test condition.}
 \label{fig:det_dvector_female}
 \end{figure}

\begin{figure}
 \centering
         \includegraphics[width=7.5cm]{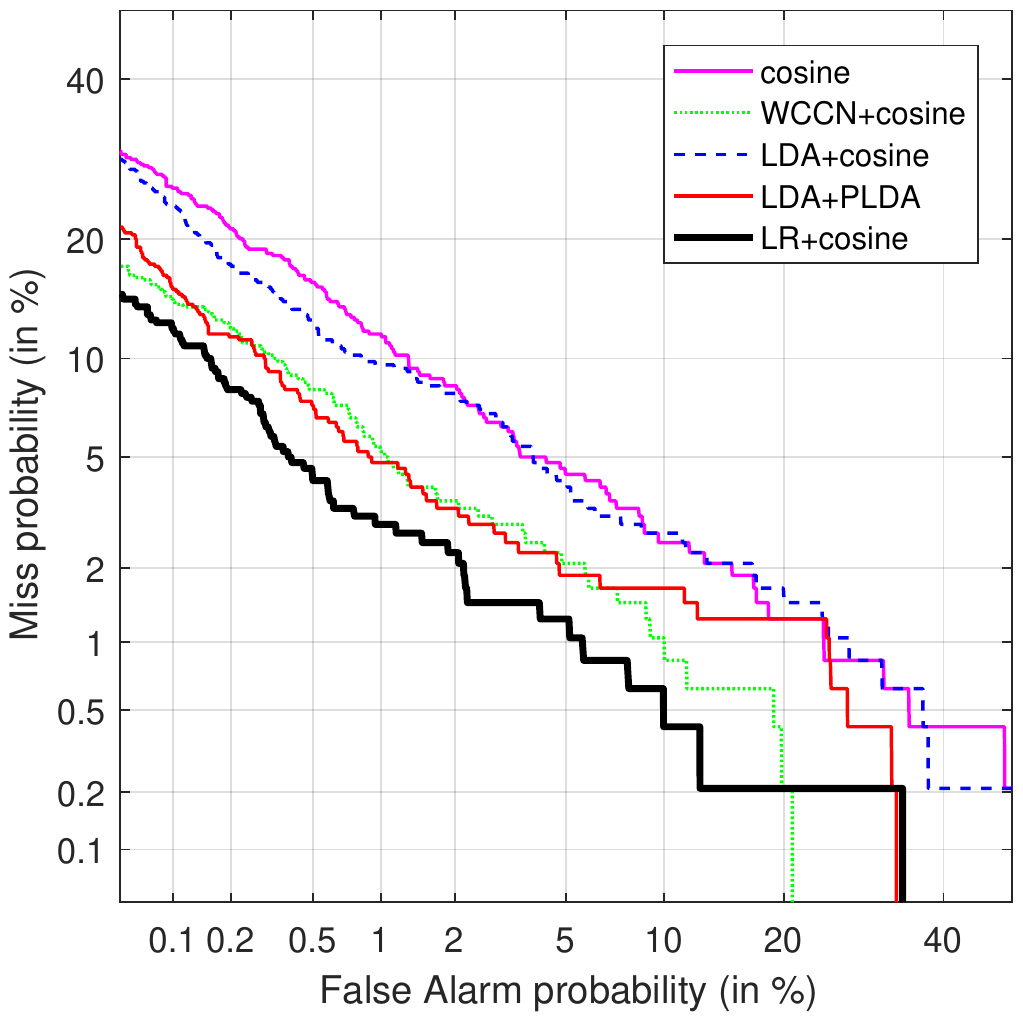}
         \caption{DET curves of the back-ends with the GMM/i-vector front-end on the male part of the initial test condition.}
 \label{fig:det_GMM_ivector_male}
 \end{figure}

\begin{figure}
 \centering
         \includegraphics[width=7.5cm]{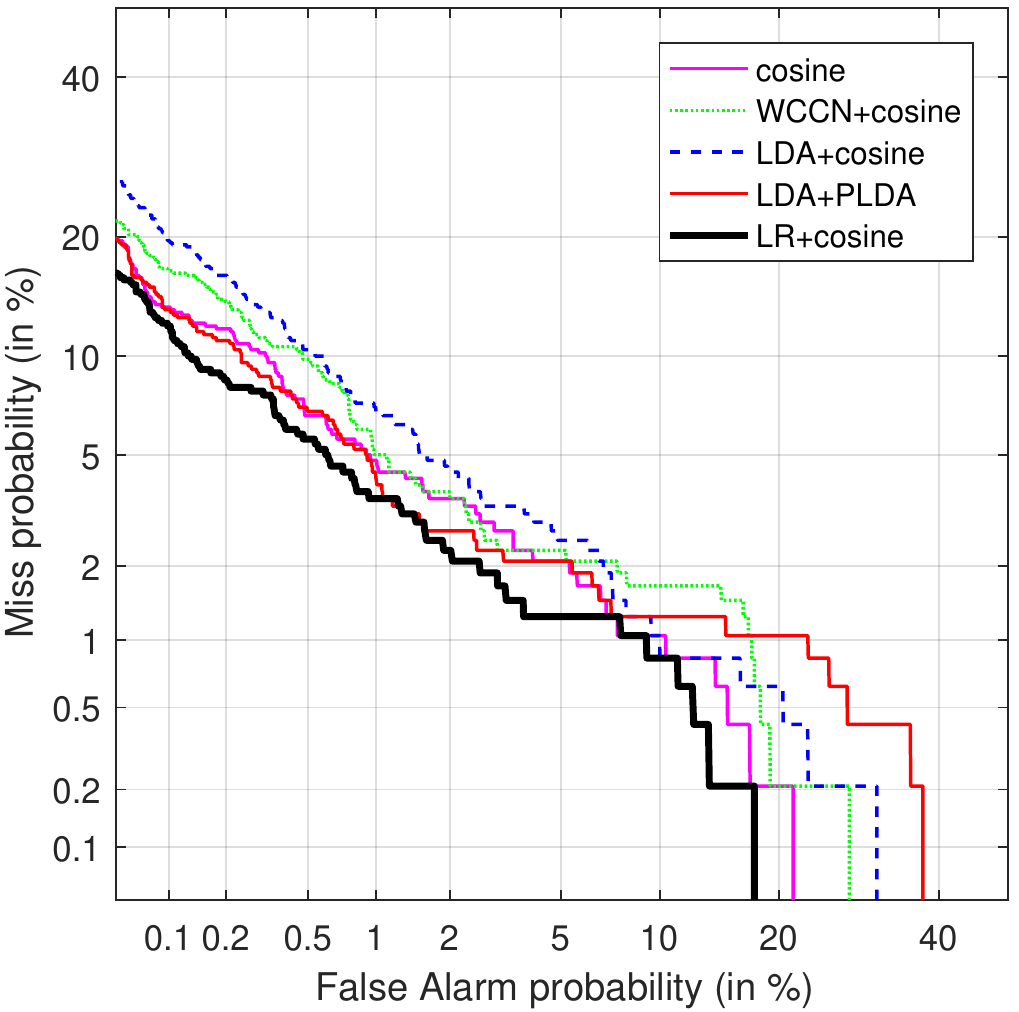}
         \caption{DET curves of the back-ends with the DNN/i-vector front-end on the male part of the initial test condition.}
 \label{fig:det_DNN_ivector_male}
 \end{figure}

\begin{figure}
 \centering
         \includegraphics[width=7.5cm]{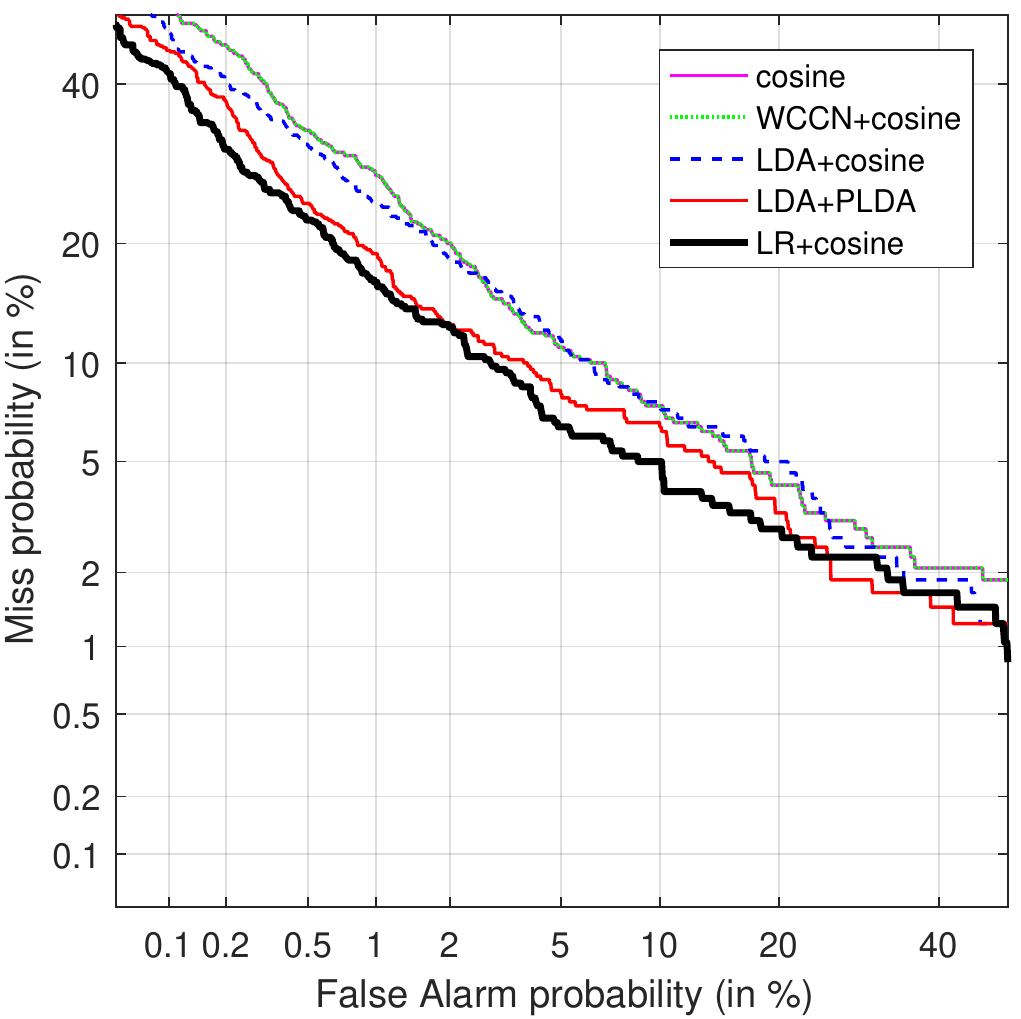}
         \caption{DET curves of the back-ends with the d-vector front-end on the male part of the initial test condition.}
 \label{fig:det_dvector_male}
 \end{figure}

We report the comparison results in the initial test condition in Figs. \ref{fig:det_GMM_ivector_female} to \ref{fig:det_dvector_male} respectively. From the figures, we observe that the proposed method outperforms the comparison methods significantly when the GMM/i-vector or DNN/i-vector front-end is used (Figs. \ref{fig:det_GMM_ivector_female}, \ref{fig:det_DNN_ivector_female}, \ref{fig:det_GMM_ivector_male} and \ref{fig:det_DNN_ivector_male}), and outperforms the comparison methods slightly when the d-vector front-end is used (Figs. \ref{fig:det_dvector_female} and \ref{fig:det_dvector_male}).

To prevent a biased conclusion that the proposed method happens to have some advantage in the initial test condition, we ran a comparison in
 the 6 test conditions described in Table \ref{tab:scenario}, where each test condition has 100 independent implementations randomly generated from the NIST 2008 SRE database. We report the average results on the male and female parts of the implementations in Tables \ref{tab:addlabel} and \ref{tab:addlabel2} respectively. From the tables, we observe that the proposed LR+cosine back-end outperforms the comparison methods when the enrollment speech is longer than 15 seconds, and is comparable to LDA+PLDA when the enrollment speech is 15 seconds long, given any of the three front-ends.

\begin{table*}[htbp]
  \centering
  \caption{Comparison results of the back-ends on the female parts of the 6 test conditions.}
  \scalebox{0.9}{
    \begin{tabular}{clccc|ccc|ccc}
\cline{3-11}          &       & \multicolumn{3}{c|}{EER (in \%)} & \multicolumn{3}{c|}{DCF$\scriptsize{\mbox{08}}$} & \multicolumn{3}{c}{DCF$\scriptsize{\mbox{10}}$} \bigstrut\\
\cline{3-11}            &       & GMM/i-vector & DNN/i-vector & d-vector & GMM/i-vector & DNN/i-vector & d-vector & GMM/i-vector & DNN/i-vector & d-vector \bigstrut\\
    \hline
    \multirow{5}[2]{*}{15"-15"} & Cosine & 16.52  & 12.14  & 9.87  & 6.2423  & 4.8058  & 4.2655  & 0.0940  & 0.0882  & 0.0873  \bigstrut[t]\\
          & WCCN+cosine & 3.78  & 4.26  & 9.87  & 1.8382  & 2.2631  & 4.2655  & 0.0625  & 0.0764  & 0.0873  \\
          & LDA+cosine & 9.70  & 9.21  & 10.03  & 3.9029  & 3.8131  & 4.0192  & 0.0805  & 0.0833  & 0.0855  \\
          & LDA+PLDA & 3.08  & \textbf{2.84} & 7.70  & 1.3656  & \textbf{1.3634} & 3.1572  & 0.0550  & \textbf{0.0575} & 0.0831  \\
          & LR+cosine & \textbf{2.80} & 3.10  & \textbf{7.45} & \textbf{1.2911} & 1.4950  & \textbf{2.9859} & \textbf{0.0528} & 0.0578  & \textbf{0.0748} \bigstrut[b]\\
    \hline
    \multirow{5}[2]{*}{30"-15"} & Cosine & 10.54  & 7.14  & 7.80  & 4.2643  & 3.0554  & 3.5023  & 0.0833  & 0.0760  & 0.0808  \bigstrut[t]\\
          & WCCN+cosine & 2.44  & 2.71  & 7.80  & 1.1682  & 1.4803  & 3.5023  & 0.0485  & 0.0629  & 0.0808  \\
          & LDA+cosine & 5.72  & 5.54  & 7.72  & 2.4234  & 2.4642  & 3.2372  & 0.0655  & 0.0706  & 0.0789  \\
          & LDA+PLDA & 2.05  & 1.85  & 5.80  & 0.9270  & 0.9054  & 2.4158  & 0.0455  & 0.0471  & 0.0750  \\
          & LR+cosine & \textbf{1.59} & \textbf{1.77} & \textbf{5.10} & \textbf{0.7326} & \textbf{0.8503} & \textbf{2.1988} & \textbf{0.0391} & \textbf{0.0435} & \textbf{0.0645} \bigstrut[b]\\
    \hline
    \multirow{5}[2]{*}{45"-15"} & Cosine & 7.63  & 5.07  & 7.02  & 3.2110  & 2.2481  & 3.2053  & 0.0745  & 0.0679  & 0.0778  \bigstrut[t]\\
          & WCCN+cosine & 1.98  & 2.22  & 7.02  & 0.9423  & 1.1902  & 3.2053  & 0.0413  & 0.0563  & 0.0778  \\
          & LDA+cosine & 4.17  & 4.12  & 6.91  & 1.7938  & 1.9061  & 2.9395  & 0.0563  & 0.0636  & 0.0752  \\
          & LDA+PLDA & 1.73  & 1.57  & 5.15  & 0.7831  & 0.7733  & 2.1448  & 0.0410  & 0.0434  & 0.0715  \\
          & LR+cosine & \textbf{1.18} & \textbf{1.34} & \textbf{4.34} & \textbf{0.5670} & \textbf{0.6585} & \textbf{1.9069} & \textbf{0.0331} & \textbf{0.0380} & \textbf{0.0592} \bigstrut[b]\\
    \hline
    \multirow{5}[2]{*}{75"-15"} & Cosine & 5.07  & 3.29  & 6.57  & 2.2101  & 1.5529  & 2.9559  & 0.0629  & 0.0586  & 0.0745  \bigstrut[t]\\
          & WCCN+cosine & 1.64  & 1.79  & 6.57  & 0.7537  & 0.9592  & 2.9559  & 0.0355  & 0.0493  & 0.0745  \\
          & LDA+cosine & 3.01  & 2.94  & 6.36  & 1.3081  & 1.4282  & 2.7023  & 0.0480  & 0.0554  & 0.0719  \\
          & LDA+PLDA & 1.53  & 1.40  & 4.62  & 0.6779  & 0.6855  & 1.9399  & 0.0376  & 0.0396  & 0.0674  \\
          & LR+cosine & \textbf{0.94} & \textbf{1.07} & \textbf{3.70} & \textbf{0.4384} & \textbf{0.5065} & \textbf{1.6927} & \textbf{0.0274} & \textbf{0.0327} & \textbf{0.0545} \bigstrut[b]\\
    \hline
    \multirow{5}[2]{*}{150"-15"} & Cosine & 2.92  & 1.89  & 6.06  & 1.2964  & 0.9553  & 2.7485  & 0.0493  & 0.0484  & 0.0713  \bigstrut[t]\\
          & WCCN+cosine & 1.35  & 1.46  & 6.06  & 0.5999  & 0.7758  & 2.7485  & 0.0308  & 0.0430  & 0.0713  \\
          & LDA+cosine & 2.10  & 2.12  & 5.81  & 0.9249  & 1.0437  & 2.4882  & 0.0402  & 0.0484  & 0.0683  \\
          & LDA+PLDA & 1.34  & 1.24  & 4.25  & 0.6010  & 0.6022  & 1.7482  & 0.0356  & 0.0371  & 0.0640  \\
          & LR+cosine & \textbf{0.74} & \textbf{0.86} & \textbf{3.25} & \textbf{0.3375} & \textbf{0.3967} & \textbf{1.4872} & \textbf{0.0233} & \textbf{0.0281} & \textbf{0.0500} \bigstrut[b]\\
    \hline
    \multirow{5}[2]{*}{225"-15"} & Cosine & 2.14  & 1.45  & 5.88  & 0.9864  & 0.7626  & 2.6622  & 0.0430  & 0.0438  & 0.0700  \bigstrut[t]\\
          & WCCN+cosine & 1.24  & 1.37  & 5.88  & 0.5564  & 0.7024  & 2.6622  & 0.0286  & 0.0409  & 0.0700  \\
          & LDA+cosine & 1.82  & 1.85  & 5.65  & 0.8148  & 0.9222  & 2.4214  & 0.0371  & 0.0449  & 0.0672  \\
          & LDA+PLDA & 1.33  & 1.22  & 4.07  & 0.5771  & 0.5913  & 1.6874  & 0.0344  & 0.0364  & 0.0619  \\
          & LR+cosine & \textbf{0.69} & \textbf{0.78} & \textbf{3.10} & \textbf{0.3155} & \textbf{0.3674} & \textbf{1.4063} & \textbf{0.0218} & \textbf{0.0265} & \textbf{0.0476} \bigstrut[b]\\
    \hline
    \end{tabular}%
    }
  \label{tab:addlabel}%
\end{table*}%

\begin{table*}[t]
  \centering
  \caption{Comparison results of the back-ends on the male parts of the 6 test conditions.}
  \scalebox{0.9}{
    \begin{tabular}{clccc|ccc|ccc}
\cline{3-11}          &       & \multicolumn{3}{c|}{EER (in \%)} & \multicolumn{3}{c|}{DCF$\scriptsize{\mbox{08}}$} & \multicolumn{3}{c}{DCF$\scriptsize{\mbox{10}}$} \bigstrut\\
\cline{3-11}          &       & \multicolumn{1}{c}{GMM/i-vector} & \multicolumn{1}{c}{DNN/i-vector} & \multicolumn{1}{c|}{d-vector} & \multicolumn{1}{c}{GMM/i-vector} & \multicolumn{1}{c}{DNN/i-vector} & \multicolumn{1}{c|}{d-vector} & \multicolumn{1}{c}{GMM/i-vector} & \multicolumn{1}{c}{DNN/i-vector} & \multicolumn{1}{c}{d-vector} \bigstrut\\
    \hline
    \multirow{5}[2]{*}{15"-15"} & Cosine & 15.74  & 7.95  & 10.16  & 5.9013  & 3.2361  & 4.0586  & 0.0898  & 0.0755  & 0.0883  \bigstrut[t]\\
          & WCCN+cosine & 4.14  & 4.13  & 10.16  & 1.8184  & 1.9061  & 4.0586  & 0.0661  & 0.0713  & 0.0883  \\
          & LDA+cosine & 9.89  & 6.55  & 10.29  & 3.8431  & 2.7328  & 3.8002  & 0.0781  & 0.0717  & 0.0785  \\
          & LDA+PLDA & 3.72  & \textbf{3.33} & 8.58  & \textbf{1.5178} & \textbf{1.4099} & 3.2521  & 0.0630  & 0.0654  & 0.0858  \\
          & LR+cosine & \textbf{3.55} & 3.44  & \textbf{8.37} & 1.5312  & 1.4960  & \textbf{3.1300} & \textbf{0.0572} & \textbf{0.0557} & \textbf{0.0793} \bigstrut[b]\\
    \hline
    \multirow{5}[2]{*}{30"-15"} & Cosine & 10.03  & 4.31  & 8.12  & 3.9863  & 1.8684  & 3.3539  & 0.0780  & 0.0598  & 0.0837  \bigstrut[t]\\
          & WCCN+cosine & 2.63  & 2.68  & 8.12  & 1.1601  & 1.2544  & 3.3539  & 0.0523  & 0.0588  & 0.0837  \\
          & LDA+cosine & 5.87  & 3.77  & 7.97  & 2.3780  & 1.6657  & 3.0572  & 0.0630  & 0.0571  & 0.0720  \\
          & LDA+PLDA & 2.50  & 2.25  & 6.46  & 1.0336  & 0.9715  & 2.5377  & 0.0544  & 0.0588  & 0.0822  \\
          & LR+cosine & \textbf{2.05} & \textbf{2.08} & \textbf{5.90} & \textbf{0.9079} & \textbf{0.9157} & \textbf{2.3789} & \textbf{0.0448} & \textbf{0.0446} & \textbf{0.0727} \bigstrut[b]\\
    \hline
    \multirow{5}[2]{*}{45"-15"} & Cosine & 7.38  & 2.94  & 7.25  & 3.0176  & 1.3420  & 3.0528  & 0.0682  & 0.0516  & 0.0810  \bigstrut[t]\\
          & WCCN+cosine & 2.08  & 2.12  & 7.25  & 0.9033  & 1.0171  & 3.0528  & 0.0451  & 0.0507  & 0.0810  \\
          & LDA+cosine & 4.29  & 2.81  & 7.00  & 1.7848  & 1.2687  & 2.7608  & 0.0545  & 0.0498  & 0.0682  \\
          & LDA+PLDA & 2.08  & 1.83  & 5.59  & 0.8581  & 0.8065  & 2.2553  & 0.0508  & 0.0550  & 0.0799  \\
          & LR+cosine & \textbf{1.54} & \textbf{1.57} & \textbf{4.99} & \textbf{0.6975} & \textbf{0.7030} & \textbf{2.0807} & \textbf{0.0385} & \textbf{0.0385} & \textbf{0.0689} \bigstrut[b]\\
    \hline
    \multirow{5}[2]{*}{75"-15"} & Cosine & 4.80  & 1.93  & 6.68  & 2.0268  & 0.9133  & 2.7933  & 0.0564  & 0.0423  & 0.0785  \bigstrut[t]\\
          & WCCN+cosine & 1.65  & 1.70  & 6.68  & 0.7231  & 0.8063  & 2.7933  & 0.0384  & 0.0431  & 0.0785  \\
          & LDA+cosine & 3.07  & 2.11  & 6.25  & 1.3144  & 0.9545  & 2.4982  & 0.0452  & 0.0417  & 0.0638  \\
          & LDA+PLDA & 1.74  & 1.56  & 5.06  & 0.7340  & 0.6900  & 1.9961  & 0.0477  & 0.0516  & 0.0778  \\
          & LR+cosine & \textbf{1.14} & \textbf{1.20} & \textbf{4.13} & \textbf{0.5295} & \textbf{0.5530} & \textbf{1.7971} & \textbf{0.0327} & \textbf{0.0323} & \textbf{0.0644} \bigstrut[b]\\
    \hline
    \multirow{5}[2]{*}{150"-15"} & Cosine & 2.92  & 1.28  & 6.28  & 1.2581  & 0.6091  & 2.6047  & 0.0442  & 0.0339  & 0.0752  \bigstrut[t]\\
          & WCCN+cosine & 1.38  & 1.42  & 6.28  & 0.6140  & 0.6831  & 2.6047  & 0.0333  & 0.0373  & 0.0752  \\
          & LDA+cosine & 2.27  & 1.61  & 5.83  & 0.9524  & 0.7351  & 2.3034  & 0.0381  & 0.0352  & 0.0606  \\
          & LDA+PLDA & 1.57  & 1.41  & 4.69  & 0.6839  & 0.6470  & 1.8220  & 0.0458  & 0.0502  & 0.0744  \\
          & LR+cosine & \textbf{0.93} & \textbf{1.03} & \textbf{3.73} & \textbf{0.4423} & \textbf{0.4682} & \textbf{1.6328} & \textbf{0.0290} & \textbf{0.0285} & \textbf{0.0602} \bigstrut[b]\\
    \hline
    \multirow{5}[2]{*}{225"-15"} & Cosine & 2.23  & 1.03  & 6.04  & 0.9779  & 0.4954  & 2.5024  & 0.0383  & 0.0302  & 0.0746  \bigstrut[t]\\
          & WCCN+cosine & 1.28  & 1.29  & 6.04  & 0.5518  & 0.6083  & 2.5024  & 0.0309  & 0.0346  & 0.0746  \\
          & LDA+cosine & 1.89  & 1.42  & 5.53  & 0.8172  & 0.6262  & 2.2010  & 0.0352  & 0.0320  & 0.0590  \\
          & LDA+PLDA & 1.49  & 1.29  & 4.41  & 0.6461  & 0.6154  & 1.7348  & 0.0453  & 0.0488  & 0.0737  \\
          & LR+cosine & \textbf{0.86} & \textbf{0.94} & \textbf{3.43} & \textbf{0.3913} & \textbf{0.4057} & \textbf{1.5194} & \textbf{0.0267} & \textbf{0.0262} & \textbf{0.0582} \bigstrut[b]\\
    \hline
    \end{tabular}%
    }
  \label{tab:addlabel2}%
\end{table*}%

\begin{figure}
 \centering
         \includegraphics[width=8.5cm]{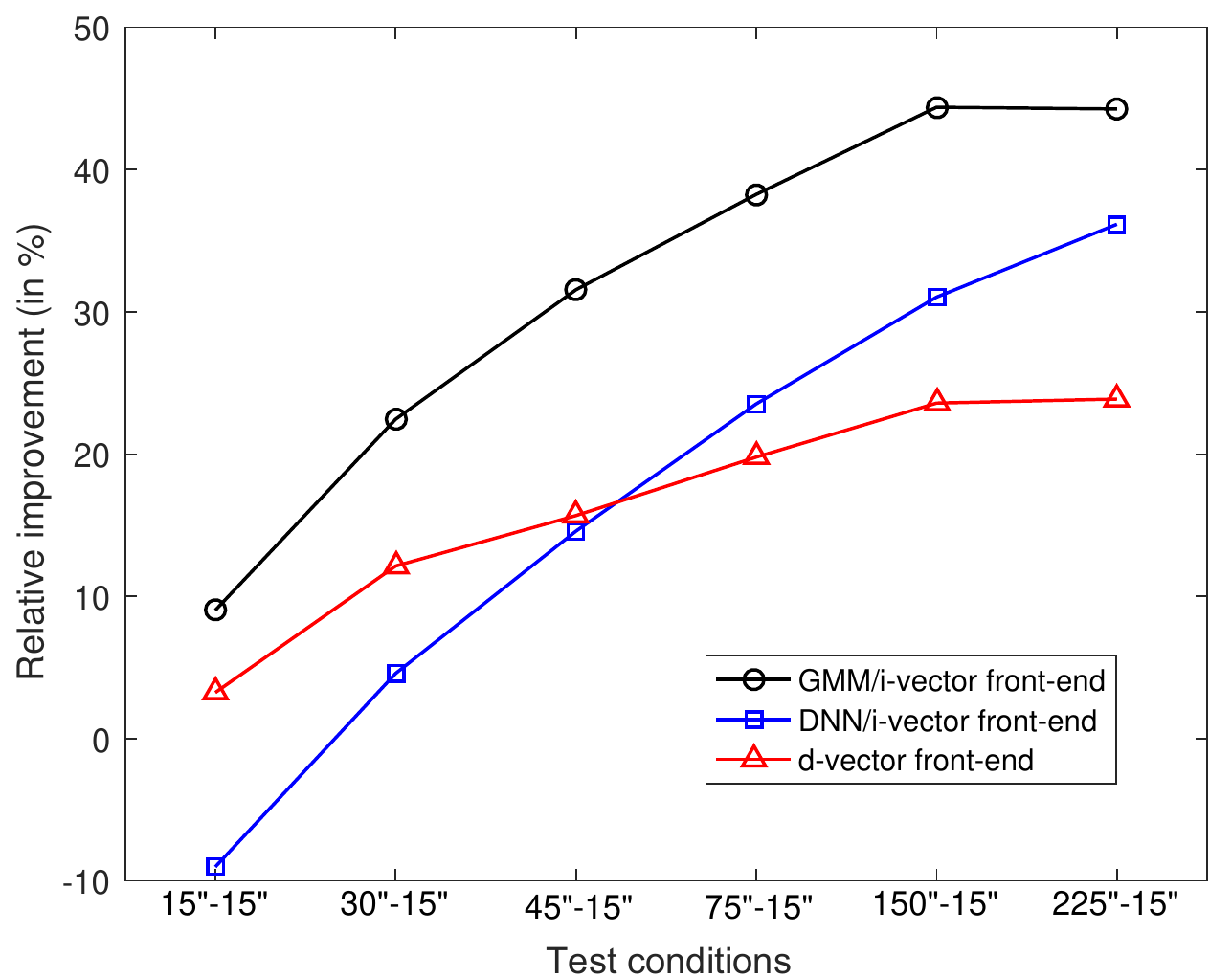}
         \caption{Relative EER improvement of the LR+cosine back-end over the best comparison methods in the female parts of the 6 test conditions.}
 \label{fig:relative_imp_female}
 \end{figure}

\begin{figure}
 \centering
         \includegraphics[width=8.5cm]{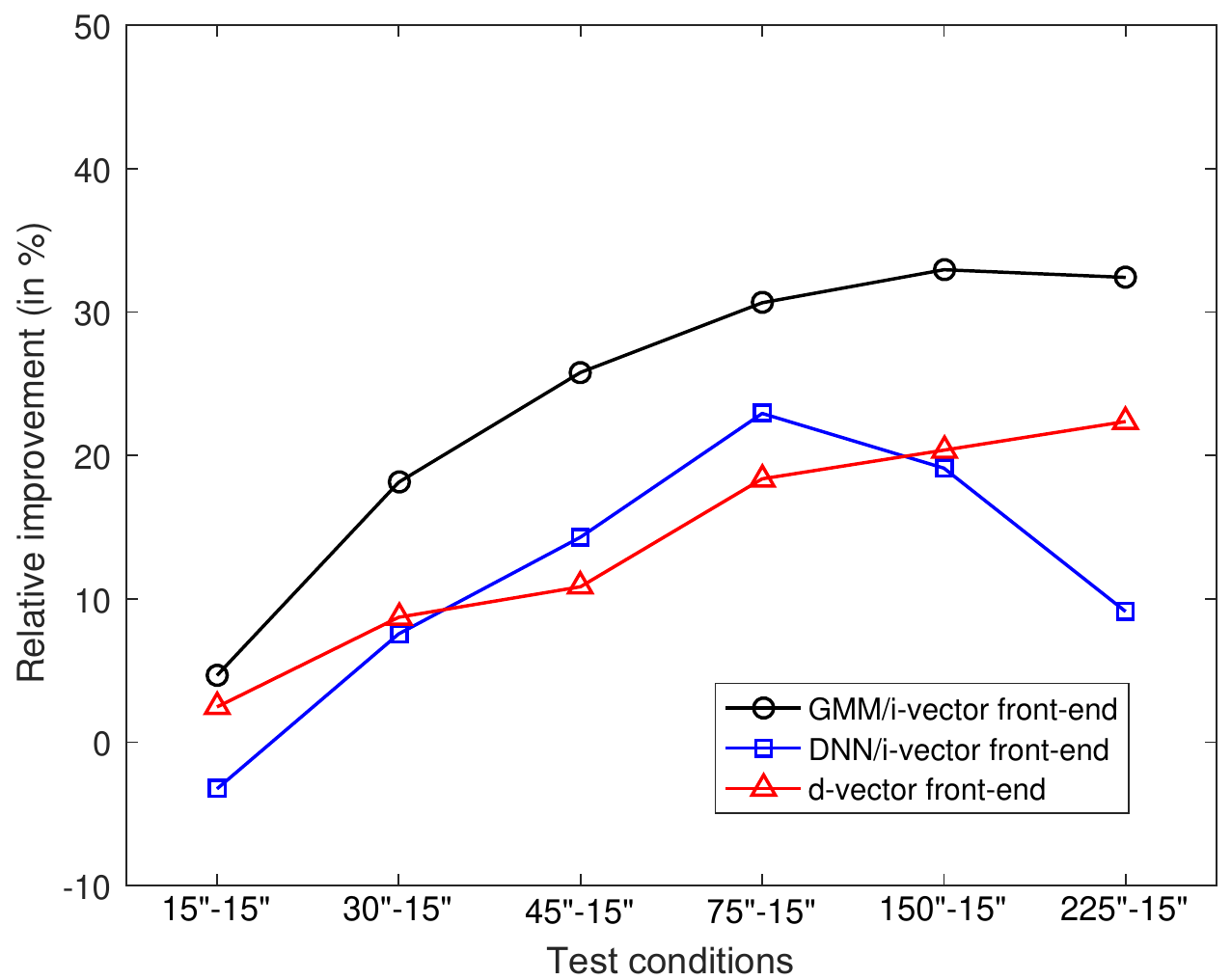}
         \caption{Relative EER improvement of the LR+cosine back-end over the best comparison methods in the male parts of the 6 test conditions.}
 \label{fig:relative_imp_male}
 \end{figure}

 We drew the curves of the relative improvement scores of the proposed method over the best comparison methods in Figs. \ref{fig:relative_imp_female} and \ref{fig:relative_imp_male}, where the relative improvement score is defined by:
     \begin{eqnarray}
\rm{Score} = \frac{\rm{EER}_{\rm{LR}}-\rm{EER}_{\rm{best\_comp}}}{\rm{EER}_{\rm{best\_comp}}}\nonumber
 \end{eqnarray}
 with $\rm{EER}_{\rm{LR}}$ and $\rm{EER}_{\rm{best\_comp}}$ denoted as the EERs of the proposed method and best comparison method respectively. From the figures, we observe the following phenomena. (i) The relative improvement is getting larger when the enrollment speech is getting longer.
 An exception is that, when the DNN/i-vector is used as the front-end, the relative improvement is not always increased for the females. This is caused by the fast performance improvement of the cosine similarity scoring when the enrollment speech is getting longer. (ii) The highest relative improvement happens with the GMM/i-vector front-end, which reaches 44.3\% for the females in the 225"-15" test condition and 33.0\% for the males in the 150"-15" test condition.

\begin{figure*}
 \centering
         \includegraphics[width=15cm]{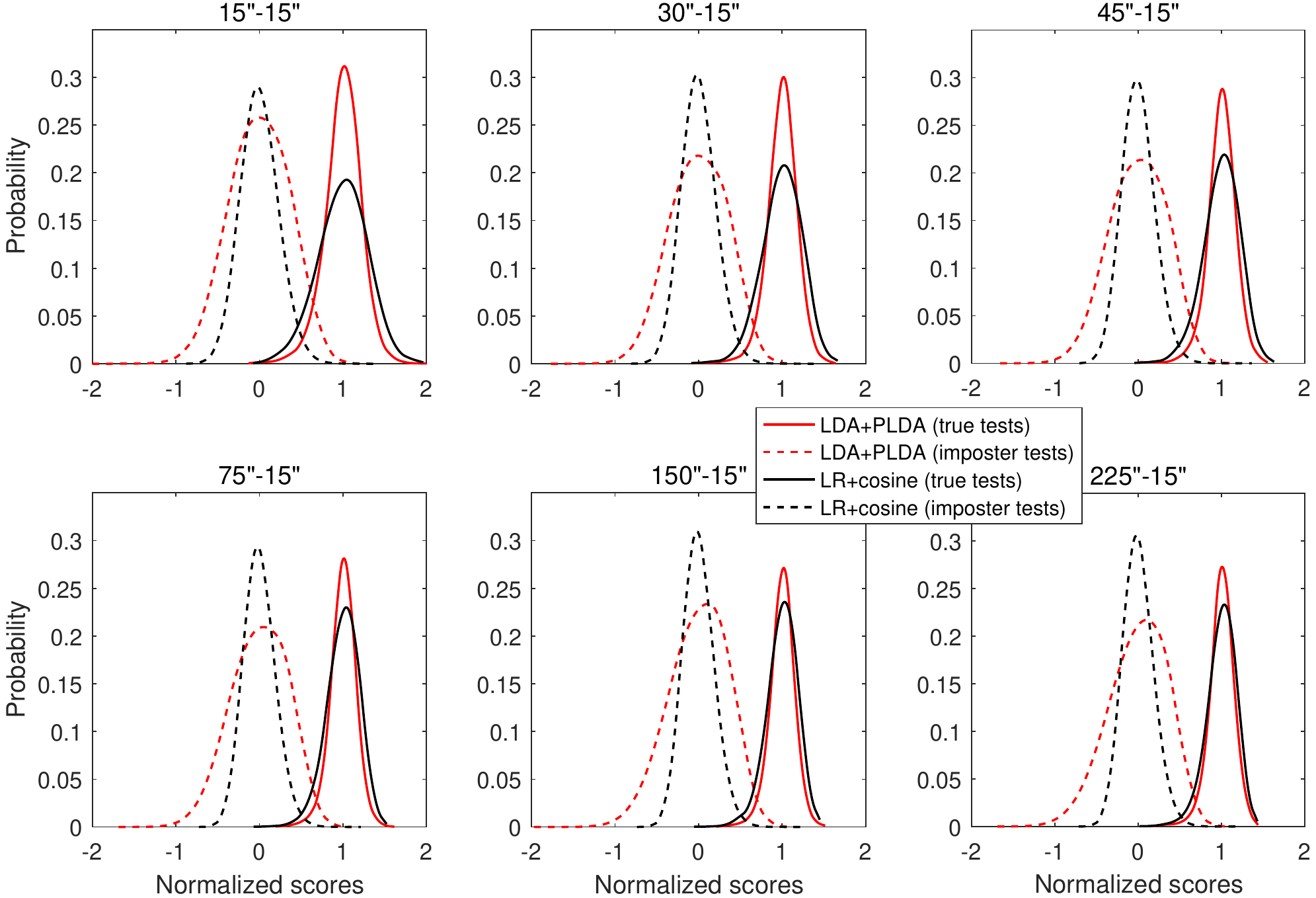}
         \caption{Histograms of the soft decision scores produced by LR+cosine and LDA+PLDA in the female parts of the 6 test conditions, where the decision scores have been normalized so that the mean values of the imposter and true trials are zero and one respectively.}
 \label{fig:hist_female}
 \end{figure*}

We also drew the soft decision scores produced from the LR+cosine and LDA+PLDA back-ends for the females in Fig. \ref{fig:hist_female} where we have normalized the decision scores to a range where the mean values of the decision scores of the imposter and true trials are zero and one respectively. From the figure, we observe that the scores produced by LR+cosine have smaller with-in class variances and smaller overlaps than those produced by LDA+PLDA.

\subsection{Effects of back-ends in fusion systems}\label{subsec:3_4}

\begin{figure}
 \centering
         \includegraphics[width=7.5cm]{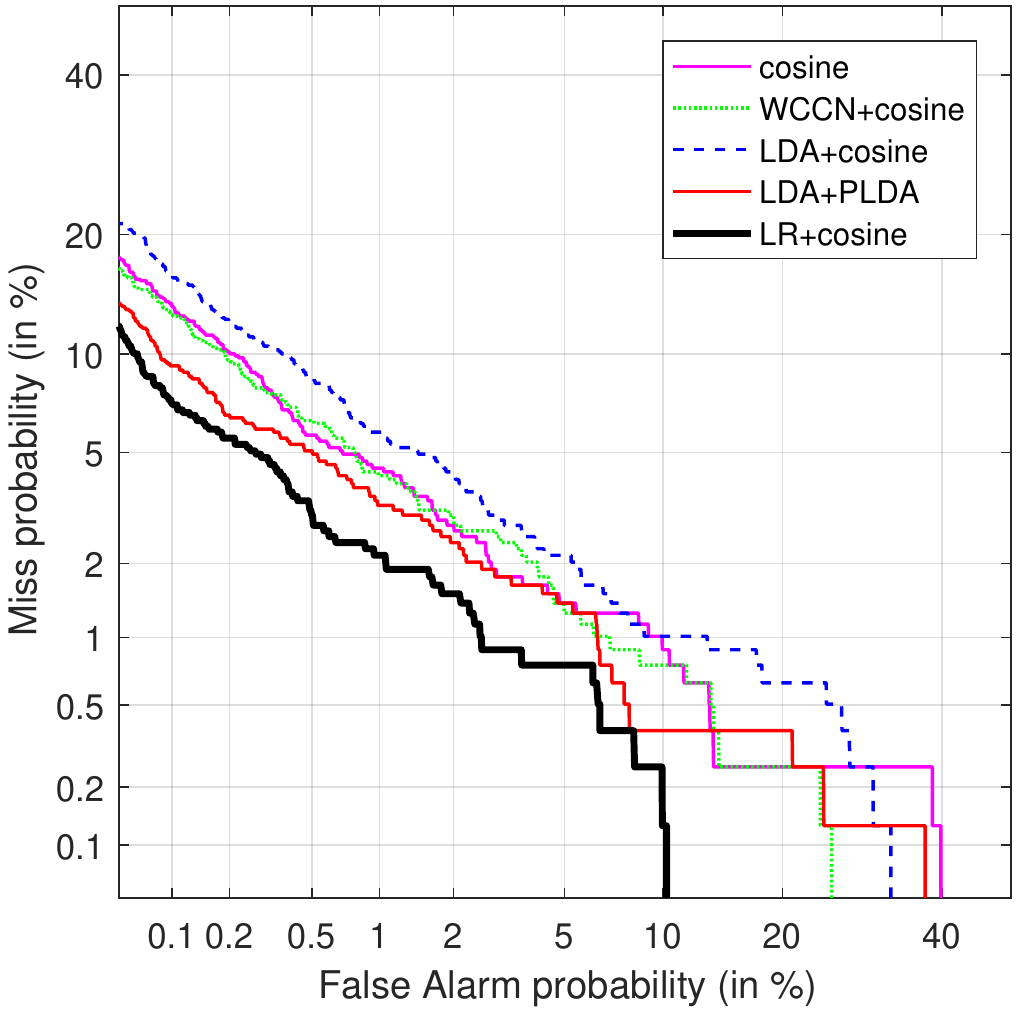}
         \caption{DET curves of the fusion systems on the female part of the initial test condition.}
 \label{fig:det_combine_female}
 \end{figure}

\begin{figure}
 \centering
         \includegraphics[width=7.5cm]{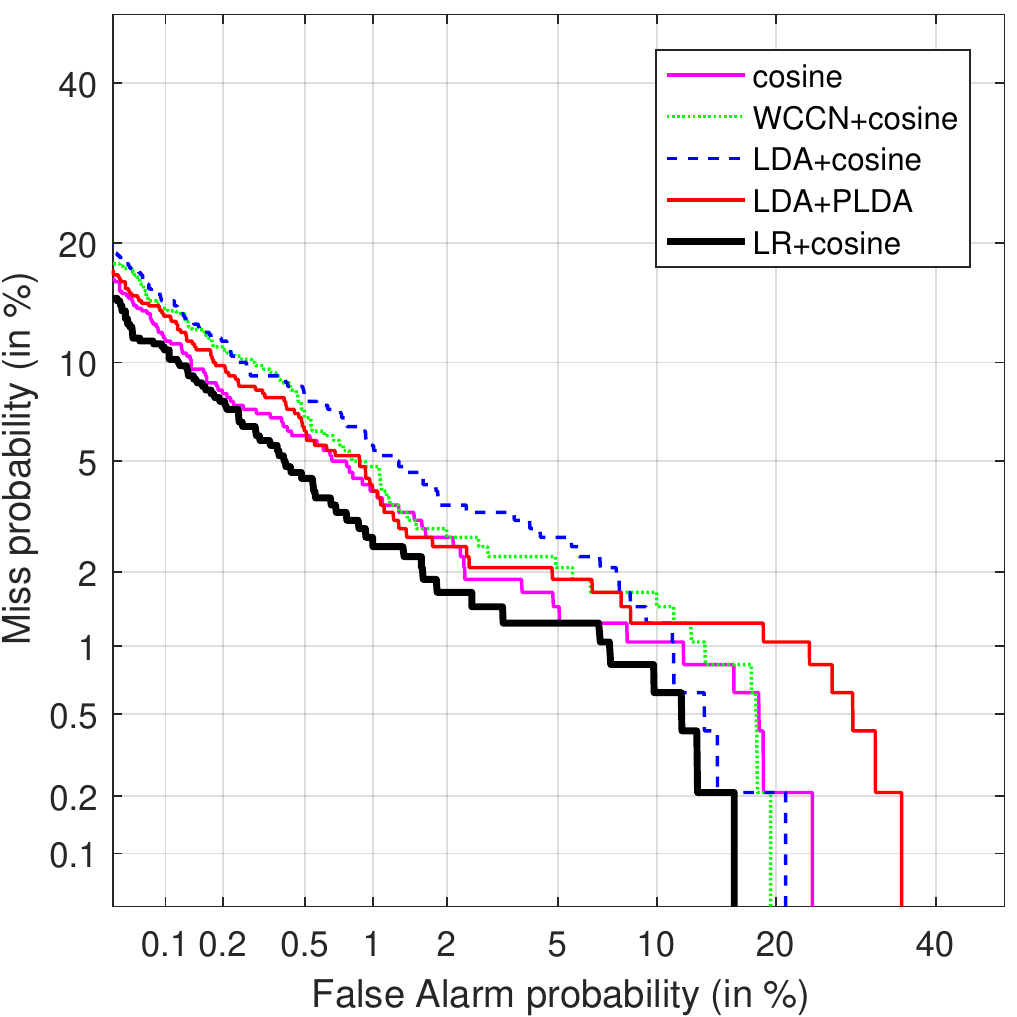}
         \caption{DET curves of the fusion systems on the male part of the initial test condition.}
 \label{fig:det_combine_male}
 \end{figure}

Fusing the decision scores produced from multiple base methods is an effective way for further improving the performance of the base methods. This subsection studies the approach of averaging the soft decision scores produced from the systems that use GMM/i-vector and DNN/i-vector as the front-ends, respectively. Figures \ref{fig:det_combine_female} and \ref{fig:det_combine_male} show the DET curves of the fusion systems with different back-ends on the initial test condition. Tables \ref{tab:addlabel3} and \ref{tab:addlabel4} list the comparison results of the fusion systems on the 6 test conditions defined in Table \ref{tab:scenario}. From the figures and tables, we observe the same experimental phenomena as those in Section \ref{subsec:result}, which supports the effectiveness of the LR+cosine back-end in the fusion systems.

Note that we have also evaluated the fusion systems that fuse the GMM/i-vector, DNN/i-vector, and d-vector front-ends together. The experimental conclusions are similar with the above.

%
%

\begin{table}[htbp]
  \centering
  \caption{Comparison results of the back-ends in the fusion systems on the female parts of the 6 test conditions.}
    \begin{tabular}{clccc}
\cline{3-5}          &       & EER (in \%) & DCF$\scriptsize{\mbox{08}}$ & DCF$\scriptsize{\mbox{10}}$ \bigstrut\\
    \hline
    \multirow{5}[2]{*}{15"-15"} & Cosine & 9.85  & 3.9613  & 0.0816  \bigstrut[t]\\
          & WCCN+cosine & 3.34  & 1.6565  & 0.0625  \\
          & LDA+cosine & 6.72  & 2.7861  & 0.0720  \\
          & LDA+PLDA & 2.64  & 1.1850  & \textbf{0.0491}  \\
          & LR+cosine & \textbf{2.46} & \textbf{1.1496} & {0.0501} \bigstrut[b]\\
    \hline
    \multirow{5}[2]{*}{30"-15"} & Cosine & 5.32  & 2.3171  & 0.0669  \bigstrut[t]\\
          & WCCN+cosine & 2.12  & 1.0516  & 0.0486  \\
          & LDA+cosine & 3.78  & 1.6648  & 0.0575  \\
          & LDA+PLDA & 1.73  & 0.7877  & 0.0399  \\
          & LR+cosine & \textbf{1.43} & \textbf{0.6557} & \textbf{0.0369} \bigstrut[b]\\
    \hline
    \multirow{5}[2]{*}{45"-15"} & Cosine & 3.59  & 1.6360  & 0.0582  \bigstrut[t]\\
          & WCCN+cosine & 1.76  & 0.8402  & 0.0416  \\
          & LDA+cosine & 2.69  & 1.2355  & 0.0498  \\
          & LDA+PLDA & 1.48  & 0.6643  & 0.0359  \\
          & LR+cosine & \textbf{1.07} & \textbf{0.5102} & \textbf{0.0317} \bigstrut[b]\\
    \hline
    \multirow{5}[2]{*}{75"-15"} & Cosine & 2.22  & 1.0728  & 0.0487  \bigstrut[t]\\
          & WCCN+cosine & 1.44  & 0.6726  & 0.0356  \\
          & LDA+cosine & 1.88  & 0.8991  & 0.0424  \\
          & LDA+PLDA & 1.31  & 0.5844  & 0.0328  \\
          & LR+cosine & \textbf{0.89} & \textbf{0.3925} & \textbf{0.0269} \bigstrut[b]\\
    \hline
    \multirow{5}[2]{*}{150"-15"} & Cosine & 1.27  & 0.6418  & 0.0393  \bigstrut[t]\\
          & WCCN+cosine & 1.18  & 0.5437  & 0.0308  \\
          & LDA+cosine & 1.35  & 0.6384  & 0.0358  \\
          & LDA+PLDA & 1.16  & 0.5038  & 0.0308  \\
          & LR+cosine & \textbf{0.72} & \textbf{0.3108} & \textbf{0.0230} \bigstrut[b]\\
    \hline
    \multirow{5}[2]{*}{225"-15"} & Cosine & 0.97  & 0.5041  & 0.0349  \bigstrut[t]\\
          & WCCN+cosine & 1.11  & 0.4976  & 0.0289  \\
          & LDA+cosine & 1.17  & 0.5593  & 0.0332  \\
          & LDA+PLDA & 1.13  & 0.4923  & 0.0300  \\
          & LR+cosine & \textbf{0.66} & \textbf{0.2911} & \textbf{0.0217} \bigstrut[b]\\
    \hline
    \end{tabular}%
  \label{tab:addlabel3}%
\end{table}%

\begin{table}[htbp]
  \centering
  \caption{Comparison results of the back-ends in the fusion systems on the male parts of the 6 test conditions.}
    \begin{tabular}{clccc}
\cline{3-5}          &       & EER (in \%) & DCF$\scriptsize{\mbox{08}}$ & DCF$\scriptsize{\mbox{10}}$ \bigstrut\\
    \hline
    \multirow{5}[2]{*}{15"-15"} & Cosine & 7.75  & 3.0791  & 0.0713  \bigstrut[t]\\
          & WCCN+cosine & 3.60  & 1.6093  & 0.0636  \\
          & LDA+cosine & 5.92  & 2.3919  & 0.0646  \\
          & LDA+PLDA & 3.21  & 1.3305  & 0.0592  \\
          & LR+cosine & \textbf{3.04} & \textbf{1.3140} & \textbf{0.0531} \bigstrut[b]\\
    \hline
    \multirow{5}[2]{*}{30"-15"} & Cosine & 4.16  & 1.7299  & 0.0557  \bigstrut[t]\\
          & WCCN+cosine & 2.36  & 1.0279  & 0.0502  \\
          & LDA+cosine & 3.23  & 1.3784  & 0.0496  \\
          & LDA+PLDA & 2.17  & 0.9102  & 0.0528  \\
          & LR+cosine & \textbf{1.78} & \textbf{0.8054} & \textbf{0.0422} \bigstrut[b]\\
    \hline
    \multirow{5}[2]{*}{45"-15"} & Cosine & 2.77  & 1.2232  & 0.0479  \bigstrut[t]\\
          & WCCN+cosine & 1.85  & 0.8134  & 0.0428  \\
          & LDA+cosine & 2.38  & 1.0190  & 0.0421  \\
          & LDA+PLDA & 1.78  & 0.7465  & 0.0486  \\
          & LR+cosine & \textbf{1.37} & \textbf{0.6244} & \textbf{0.0358} \bigstrut[b]\\
    \hline
    \multirow{5}[2]{*}{75"-15"} & Cosine & 1.71  & 0.7997  & 0.0388  \bigstrut[t]\\
          & WCCN+cosine & 1.46  & 0.6495  & 0.0359  \\
          & LDA+cosine & 1.68  & 0.7529  & 0.0346  \\
          & LDA+PLDA & 1.51  & 0.6403  & 0.0456  \\
          & LR+cosine & \textbf{1.04} & \textbf{0.4770} & \textbf{0.0304} \bigstrut[b]\\
    \hline
    \multirow{5}[2]{*}{150"-15"} & Cosine & 1.08  & 0.5223  & 0.0309  \bigstrut[t]\\
          & WCCN+cosine & 1.27  & 0.5602  & 0.0313  \\
          & LDA+cosine & 1.26  & 0.5737  & 0.0295  \\
          & LDA+PLDA & 1.34  & 0.5974  & 0.0442  \\
          & LR+cosine & \textbf{0.89} & \textbf{0.4074} & \textbf{0.0269} \bigstrut[b]\\
    \hline
    \multirow{5}[2]{*}{225"-15"} & Cosine & 0.85  & 0.4194  & 0.0274  \bigstrut[t]\\
          & WCCN+cosine & 1.12  & 0.4976  & 0.0286  \\
          & LDA+cosine & 1.06  & 0.4904  & 0.0267  \\
          & LDA+PLDA & 1.27  & 0.5681  & 0.0429  \\
          & LR+cosine & \textbf{0.80} & \textbf{0.3586} & \textbf{0.0247} \bigstrut[b]\\
    \hline
    \end{tabular}%
  \label{tab:addlabel4}%
\end{table}%

\section{Conclusions}
\label{sec:conclusion}

In this paper, we have presented a speaker verification back-end based on linear regression. Linear regression is a simple linear model that minimizes the mean squared estimation error between the target and its estimate with a closed form solution, where the target for our speaker verification problem is defined as the ground-truth indicator vectors of utterances. The proposed LR+cosine back-end first learns speaker models by the LR model, and then applies the cosine similarity scoring to evaluate the similarity of a pair of speaker models. We have further proposed three LR-based speaker verification systems by combining the LR+cosine back-end with the GMM/i-vector, DNN/i-vector, and d-vector front-ends respectively. We have conducted an extensive experiment on the NIST 2006 SRE and NIST 2008 SRE data sets, where we used the 8\textit{conv} condition of the NIST 2006 SRE for development and the 8\textit{conv} condition of the NIST 2008 SRE for enrollment and test. To prevent a biased experimental conclusion on a particular evaluation environment, the experiment was carried out with different lengths of enrollment speech covering a range from 15 seconds to 225 seconds and repeated 100 times. The experimental results show that the proposed LR+cosine back-end outperforms several common back-ends including the cosine, WCCN+cosine, LDA+cosine, and LDA+PLDA back-ends in most cases in terms of DET curves, EER, DCF$\scriptsize{\mbox{08}}$, and DCF$\scriptsize{\mbox{10}}$.

\section{Acknowledgements}
\label{sec:prior}

This work was supported in part by the Natural Science Foundation of China under Grant No. 61671381.

\bibliographystyle{IEEEtran}
\bibliography{zxlrefs,zxlrefs3,mywork}

\end{document}